\documentclass[11pt]{article}
\usepackage{hyperref}
\pdfoutput=1
\begin{document}
\title{Flying in Two Dimensions}
\author{Manu Prakash and Donald Kim\\  
\\ Stanford University, Stanford, CA 94305, USA}
\maketitle


Diversity and specialization of behavior in insects is unmatched.  Insects hop, walk, run, jump, row, swim, glide and fly to propel themselves in a variety of environments. We have uncovered an unusual mode of propulsion of aerodynamic flight in two dimensions in adult Waterlilly Beetles \emph{(Galerucella)}. The adult beetles, often found in water lilly ponds, propel themselves strictly in a two-dimensional plane on the surface of water via flapping wing flight. Here we analyze the aerodynamics of this peculiar flight mode with respect to forces exerted on the organism during flight. The complexity of 2-D flight is captured by accounting for additional forces beyond gravitational, thrust, lift and drag, exerted on the insect body in 3D flight. Understanding this constrained propulsion mode requires accounting for viscous drag, surface tension, buoyancy force, and capillary-gravity wave drag. Moreover, dramatic differences exist in the magnitude of the resultant forces and wing strokes for sustained 2D vs. 3D flight. Here, in this fluid dynamics video, we discuss this unusual 2D flight mode via kinematic and non-dimensional analysis comparing it to 3D flight and draw generalized lessons towards understanding surface skimming based origin of flight in insects. 

Several hundred species of adult waterlilly beetles (5-6mm in body length) are found throughout North America. We have uncovered several modes of flight displayed by adult beetles including pure two dimensional flight on the fluid interface (linear horizontal velocity 30cm/sec), pure three dimensional flight and a transition from 2D flight on the water surface to free 3D flight (take-off). All 2D modes display a permanent contact of the legs with the water surface with either two or four contact points. This also serves as the site for additional applied forces on the leg such as viscous drag, surface tension force, buoyancy force, and capillary-wave drag.

Beetles were captured during summer at Harvard Research Forest, MA. Here, in order to analyze the relative contribution of the forces on the leg, we made experimental observation on two different species of water lilly beetles using a high-speed camera (Phantom v310 CMOS digital camera, Vision Research) with the frame rate of 3000 fps. The videos were captured in lab settings and all modes described above were discovered in our video data. Various parameters distinguishing 2D and 3D modes were extracted from the data, including wing stroke, translational velocity and body angle with respect to the water surface, further utilized to mark the transition from 2D to 3D flight mode. 

Finally, by using the velocity data, we analyzed the degree of contribution of the exerted forces in water-leg interface using non-dimensional analysis. With reasonable assumptions on the size of the beetles and drag coefficients, we evaluate the range of Reynolds number, Froude number and Weber number during 2D and 3D flight.  Several assumptions in the following calculations include: 1) we considered the body and leg of insect are stiff and do not change the geometry so that the velocity of insect is identical in every reference point. 2) even though the real leg is composed of many tiny hairs in microscopic point of view, we simplified the figure as cylindrical shape. From these presumptions, we could get the results at the same time frames with the velocity analysis. By putting inertia as a standard value (= 1.0), we could further compare the relative magnitude of gravity, surface tension, and viscosity. Throughout the 2D flight, the degrees of factors are Surface Tension $>>$ Inertia $>$ Gravity $>$ Viscosity.

Surface-skimming on water in stoneflies has been proposed as a mode that led to the origin of flight in insects, allowing flight control, wing morphology and wing musculature to evolve without requiring significant lift generation. Our observations provide the first evidence of a surface-skimming mode outside stoneflies.  

\end{document}